\RequirePackage{snapshot}
\pdfoutput=1
\documentclass[a4paper,11pt]{article}

\usepackage{url}

\usepackage{fullpage}

\usepackage{amsmath,amsthm,amssymb}
\usepackage{mathabx}
\usepackage{xcolor}
\usepackage{graphicx}
\usepackage{bbold}
\usepackage{textcomp} 
\usepackage{siunitx}

\usepackage{authblk}
\renewcommand\Affilfont{\itshape\small}

\renewcommand{\vec}[2][]{\boldsymbol{#2}_\text{#1}}

\newcommand{\vnabla}{\vec{\nabla}}
\newcommand{\dx}{\;\text{d}x}

\def\lambdasf{\lambda_\text{sf}}
\def\lambdaj{\lambda_\text{J}}
\def\tausf{\tau_\text{sf}}
\def\je{\vec[e]{j}}
\def\js{\vec[s]{j}}
\def\Ms{M_\text{s}}
\def\heff{\vec[eff]{h}}
\def\seff{s_\text{eff}}

\title{
  Field- and damping-like spin-transfer torque in magnetic multilayers
}

\author[1]{Claas Abert}
\author[2]{Hossein Sepehri-Amin}
\author[1]{Florian Bruckner}
\author[3]{Christoph Vogler}
\author[2,4]{Masamitsu Hayashi}
\author[1]{Dieter Suess}
\affil[1]{Christian Doppler Laboratory of Advanced Magnetic Sensing and Materials, Institute of Solid State Physics, TU Wien, Austria}
\affil[2]{National Institute for Materials Science, Tsukuba 305-0047, Japan}
\affil[3]{Institute of Solid State Physics, TU Wien, Austria}
\affil[4]{Department of Physics, The University of Tokyo, Bunkyo, Tokyo 113-0033, Japan}

\begin{document}

\maketitle
\begin{abstract}
  We investigate the spin-transfer torque in a magnetic multilayer structure by means of a spin-diffusion model.
  The torque in the considered system, consisting of two magnetic layers separated by a conducting layer, is caused by a perpendicular-to-plane current.
  We compute the strength of the field-like and the damping-like torque for different material parameters and geometries.
  Our studies suggest that the field-like torque highly depends on the exchange coupling strength of the itinerant electrons with the magnetization both in the pinned and the free layer.
  While a low coupling leads to very high field-like torques, a high coupling leads to low or even negative field-like torques.
  The dependence of the different torque terms on system parameters is considered very important for the development of applications such as STT MRAM and spin-torque oscillators.
\end{abstract}
\newpage

\section{Introduction}
Recently proposed magnetic storage technologies exploit the interaction of spin-polarized currents with the magnetization due to spin torque.
Prominent examples for such devices are spin-transfer torque magnetic random-access memories (STT MRAM) \cite{huai2008spin,lin200945nm,khvalkovskiy2013basic} and spin-torque oscillators (STO) that serve as field generators for microwave assisted recording of hard-disk drives \cite{zhu2008microwave,zhu2010microwave}.

It is understood that the origin of spin torque is the interaction of spin-polarized conducting electrons with localized magnetic moments \cite{ralph2008spin}.
In semiclassical theories, such as micromagnetics, this polarization is represented by the spin accumulation $\vec{s}$ which describes the deviation of the spin carried by conducting electrons in the presence of charge current $\je > 0$ from the equilibrium situation at $\je = 0$.
Magnetization dynamics under the influence of an effective field $\heff$ and spin accumulation $\vec{s}$ is governed by the Landau-Lifshitz-Gilbert equation (LLG)
\begin{equation}
  \frac{\partial \vec{m}}{\partial t} = 
  -\gamma \vec{m} \times \left( \heff + \frac{J}{\hbar \gamma \Ms} \vec{s} \right)
  +\alpha \vec{m} \times \frac{\partial \vec{m}}{\partial t}
  \label{eq:llg}
\end{equation}
where $\vec{m}$ is the normalized magnetization, $\gamma$ is the gyromagnetic ratio, $\alpha$ is the Gilbert damping, and $J$ is the exchange strength between conducting electrons and magnetization.

A well-known system that exploits spin-torque effects is a three-layer structure consisting of two magnetic layers separated by a conducting nonmagnetic spacer layer.
When applying a charge current perpendicular to the layers, one of the magnetic layers, referred to as pinned layer, acts as a spin polarizer.
When the spin polarized electrons reach the second layer, referred to as free layer, they accumulate at the interface and thereby exert a torque onto the free layer.
The pinned layer generates a current with a spin polarization $\vec{M}$ parallel to its magnetization $\vec[pinned]{m}$.
Hence, it is natural to investigate the torque with respect to the polarization $\vec{M}$.
The spin diffusion $\vec{s}$ can be written in a basis constructed by the magnetization $\vec{m}$ and a reference polarization $\vec{M}$
\begin{equation}
  \vec{s} =
  a \vec{M} \times \vec{m} + b (\vec{m} \times \vec{M}) \times \vec{m} + c \vec{m}.
\end{equation}
Inserting into \eqref{eq:llg} and considering $\|\vec{m}\| = 1$ yields
\begin{align}
  \frac{\partial \vec{m}}{\partial t} = 
  -\gamma \vec{m} \times \left( \heff + \frac{Jb}{\hbar \gamma \Ms} \vec{M} \right)
  -\vec{m} \times \left(\frac{Ja}{\hbar \Ms} \vec{M} \times \vec{m} \right)
  +\alpha \vec{m} \times \frac{\partial \vec{m}}{\partial t}.
  \label{eq:llg_with_s}
\end{align}
The torque term added to the effective field $\heff$ is usually referred to as field-like torque \cite{tulapurkar2005spin}.
The second term is called spin-transfer torque or damping-like torque, since it essentially leads to a relaxation of the magnetization $\vec{m}$ in direction of the polarization $\vec{M}$.
Accordingly, the coefficients $a$ and $b$ are proportional to the strength of the damping-like and field-like torque respectively.

Despite its naming, the damping-like torque does also contribute to the precessional motion of the magnetization.
Also, the field-like torque leads to both precessional and damping-like motion of the magnetization.
This can be seen by transforming \eqref{eq:llg_with_s} into the explicit form of the LLG that reads
\begin{equation}
  \frac{\partial \vec{m}}{\partial t} =
  \frac{J}{\hbar \Ms (1 + \alpha^2)} \big[
    - (b + \alpha a)  \vec{m} \times \vec{M}
    - (-a + \alpha b) \vec{m} \times (\vec{m} \times \vec{M})
  \big].
  \label{eq:llg_explicit_with_s}
\end{equation}
In materials with low damping $\alpha \ll 1$, the field-like torque and the damping-like torque, as previously defined, can be identified with precessional and damping-like motion of the magnetization respectively.
However, in materials with high damping $\alpha \approx 1$ there is a strong intermixing of the field-like and damping-like contributions.

A simple model for the description of spin torque in multi-layer structures is the macro-spin model by Slonczewski \cite{slonczewski1996current}.
In this model the coefficients $a$ and $b$ are defined in terms of the angle between porlarization $\vec{M}$ and the magnetization in the free layer $\vec{m}$ as well as some general constants that depend on material parameters and the geometry of the system.
However, the exact nature of this dependency is not provided by the model and thus the model constants are usually obtained by fitting simulation results to experimental data.

In this work we use a drift-diffusion model to compute the spin accumulation $\vec{s}$ directly from material parameters and geometry.
By projection onto the basis functions $\vec{M} \times \vec{m}$ and $(\vec{m} \times \vec{M}) \times \vec{m}$ we obtain the spatially resolved coefficients $a$ and $b$.

\section{Model}
According to the drift-diffusion model introduced in \cite{zhang2002mechanisms} the spin accumulation $\vec{s}$ is defined as
\begin{equation}
  \frac{\partial \vec{s}}{\partial t}
  = - \vnabla \cdot \js - \frac{\vec{s}}{\tausf} - J \frac{\vec{s} \times \vec{m}}{\hbar} 
  \label{eq:spin_accumulation}
\end{equation}
where $\js$ is the spin current, $\tausf$ is the spin-flip relaxation time and $J$ is the same coupling constant as in $\eqref{eq:llg}$.
The spin current is given by
\begin{equation}
  \js =
  \beta \frac{\mu_\text{B}}{e} \vec{m} \otimes \je
  - 2 D_0 \left[
    \vnabla\vec{s}
    - \beta \beta' \vec{m} \otimes ( (\vnabla\vec{s})^T \vec{m})
  \right]
  \label{eq:spin_current}
\end{equation}
where $\beta$ and $\beta'$ are dimensionless polarization parameters and $D_0$ is the diffusion constant.
Instead of the spin-flip relaxation time $\tausf$ and the coupling constant $J$, the material is often described in terms of the characteristic lengths $\lambdasf = \sqrt{2 D_0 \tausf}$ and $\lambdaj = \sqrt{2 D_0 \hbar / J}$.
Throughout this work we will use these material parameters that are more common in the experimental community.
We numerically solve \eqref{eq:spin_accumulation} assuming equilibrium $\partial \vec{s} / \partial t = 0$ with the finite-element method along the lines of \cite{ruggeri2016coupling}.
Since the spin accumulation relaxes two orders of magnitude faster than the magnetization, this assumption has no significant influence on the magnetization dynamics.
We apply homogeneous Neumann boundary conditions for $\vec{s}$ which corresponds to vanishing spin current at the boundaries, see \cite{abert2015self}.
In experiments the multi-layer structure is usually contacted with nonmagnetic leads whose thicknesses are well above the decay length of the spin accumulation which justifies the no-flux boundary condition.
However, for the numerical solution of \eqref{eq:spin_accumulation} this would require a huge amount of finite elements to discretize the leads.
In order to avoid computational overhead we use effective material parameters that allow us to model infinite leads with a thin nonmagnetic layer, see appendix~\ref{sec:effective_leads}.

\begin{figure}
  \centering
  \includegraphics{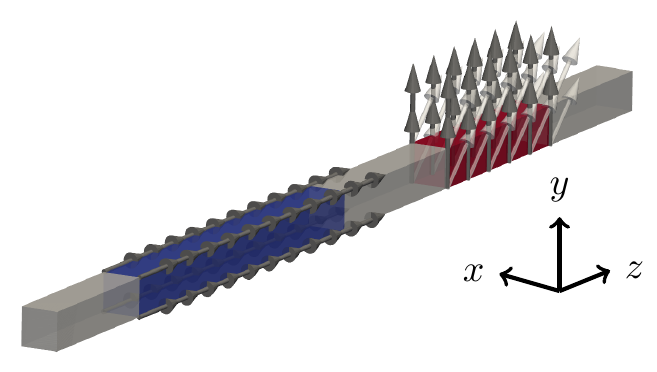}
  \caption{
    The quasi one-dimensional model. The fixed layer marked in blue is homogeneously magnetized in $z$-direction. The free layer marked in red is homogeneously magnetized in the $yz$-plane.
    Two different configurations of the free layer are depicted.
  }
  \label{fig:model}
\end{figure}
In order to investigate the influence of material parameters and geometry onto the different torque terms, we consider the quasi one-dimensional system depicted in Fig.~\ref{fig:model}.
The system consists of two magnetic layers, a pinned layer (\SI{10}{nm}) and a free layer (\SI{5}{nm}).
It is completed with a nonmagnetic spacer layer (\SI{5}{nm}) and two nonmagnetic leads (\SI{4}{nm}).
For the magnetic layers, we choose material parameters similar to those of Heusler alloys, namely $D_0 = \SI{1e-3}{m^2/s}$, $\beta = \beta' = 0.8$, $\lambdasf = \SI{8}{nm}$, and $\lambdaj = \SI{1}{nm}$ \cite{nakatani2010bulk}.
For the spacer we choose parameters similiar to Ag ($D_0 = \SI{5e-3}{m^2/s}$ and $\lambdasf = \SI{100}{nm}$) and for the leads we choose parameters similar to Au ($D_0 = \SI{5e-3}{m^2/s}$ and $\lambdasf = \SI{35}{nm}$) \cite{bass2016cpp}.
For homogeneous magnetization configurations, as considered in this work, the lateral dimension of the system does not have any impact on the solution of the spin accumulation and spin torque.
Hence, we choose very small lateral dimensions (quasi one-dimensional) in order to speed up computations.
Note, that in order to simulate infinite leads, we compute an effective diffusion length according to \eqref{eq:effective_lambdasf} which leads to $\lambdasf^\text{eff} \approx \SI{11.6}{nm}$ for a lead width of \SI{4}{nm}.

\section{Results}
The spin accumulation for the multilayer stack described in the preceding section is computed for a constant current $j_\text{e} = \SI{e12}{A/m^2}$ flowing perpendicular to the layers.
Note that we compute the torques for a given magnetization configuration without considering the resulting dynamics of the magnetization.
Since the spin accumulation, and thus also the torques, scale linearly with the current strength $j_\text{e}$, the choice of $j_\text{e}$ does not have any influence on the qualitative results of this work.
The current direction is chosen such that the conducting electrons pass the pinned layer before entering the free layer.
The magnetization in the pinned layer (and thus also $\vec{M}$) is set homogeneously in $z$-direction, perpendicular to the layers.
Unless specified differently, the magnetization in the free layer is set homogeneously in $y$-direction and the geometry as well as the material parameters are chosen according to the preceding section.
The resulting spin accumulation $\vec{s}$ is projected onto $\vec{M} \times \vec{m}$ and $(\vec{m} \times \vec{M}) \times \vec{m}$ respectively to obtain the strength of the damping-like torque $T_\text{damp} = -Ja/\hbar \gamma \Ms$ and the strength of the field-like torque $T_\text{field} = Jb/\hbar \gamma \Ms$.
Averaging over the free layer results in $\langle T_\text{damp} \rangle = \SI{42444}{A/m}$ and $\langle T_\text{field} \rangle = \SI{2712}{A/m}$.
These values are in good accordance with experimental findings \cite{zimmler2004current}.
The torques have a positive sign, i.e. the rotation caused by the field-like torque has the same direction as the rotation caused by an external field directed in the orientation of the pinned layer and the damping-like torque is directed towards the orientation of the pinned layer.
Moreover, the damping-like torque is an order of magnitude larger than the field-like torque.

\begin{figure}
  \includegraphics{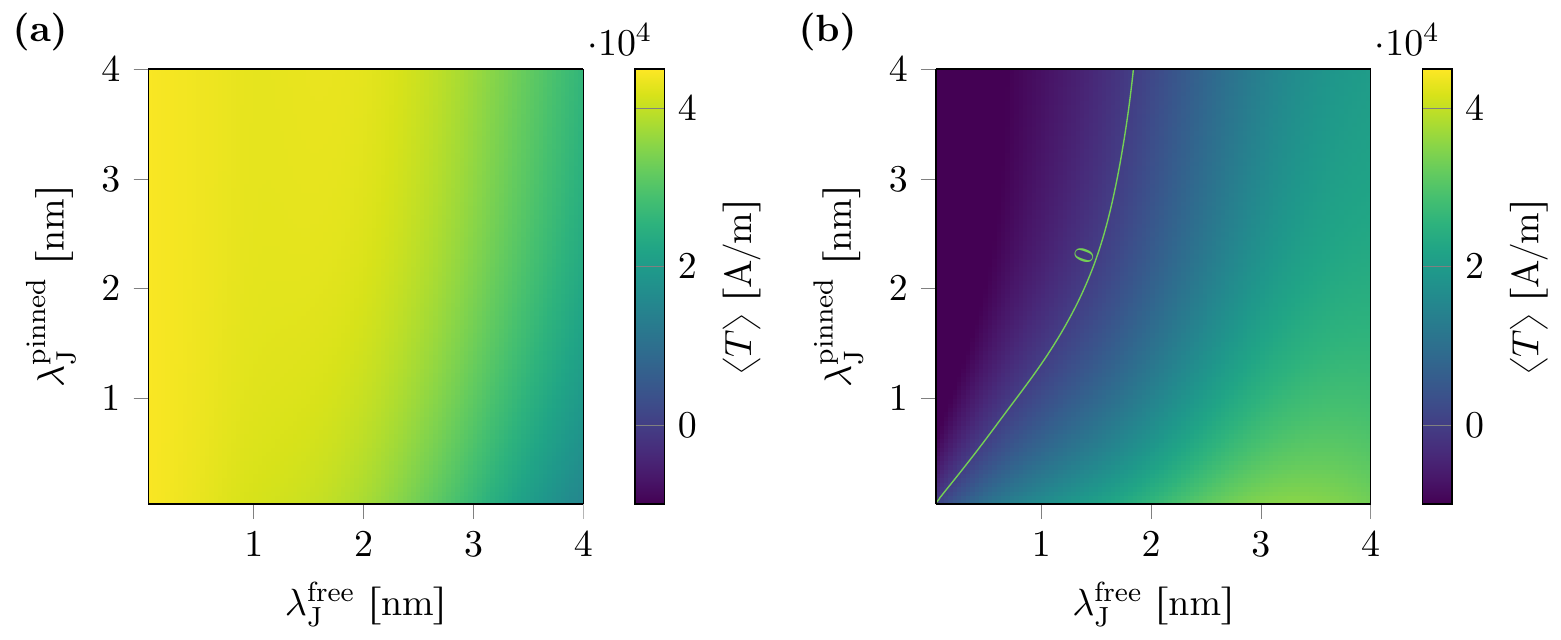}
  \caption{
    Damping-like and field-like torque for different exchange lengths in the pinned layer $\lambdaj^\text{pinned}$ and the free layer $\lambdaj^\text{free}$.
    (a) Damping-like torque $T_\text{damp}$.
    (b) Field-like torque $T_\text{field}$.
  }
  \label{fig:ljp_ljf}
\end{figure}
In a next experiment the influence of the exchange coupling of itinerant electrons and magnetization onto the different torque terms is investigated.
Fig.~\ref{fig:ljp_ljf} shows the resulting $\langle T_\text{damp} \rangle$ and $\langle T_\text{field} \rangle$ for varying $\lambdaj$ both in the free layer and the pinned layer.
We choose $\lambdaj$ as $0-\SI{4}{nm}$, which is a realistic range according to \cite{shpiro2003self}.
The strength of the damping-like torque does not significantly depend on the choice of $\lambdaj$, see Fig.~\ref{fig:ljp_ljf}(a), 
However, the field-like torque strength shows a very pronounced dependency.
As shown in Fig.~\ref{fig:ljp_ljf}(b) not only the strength, but also the sign of the field-like torque may change depending on the choice of $\lambdaj$ in the pinned and free layer.

\begin{figure}
  \includegraphics{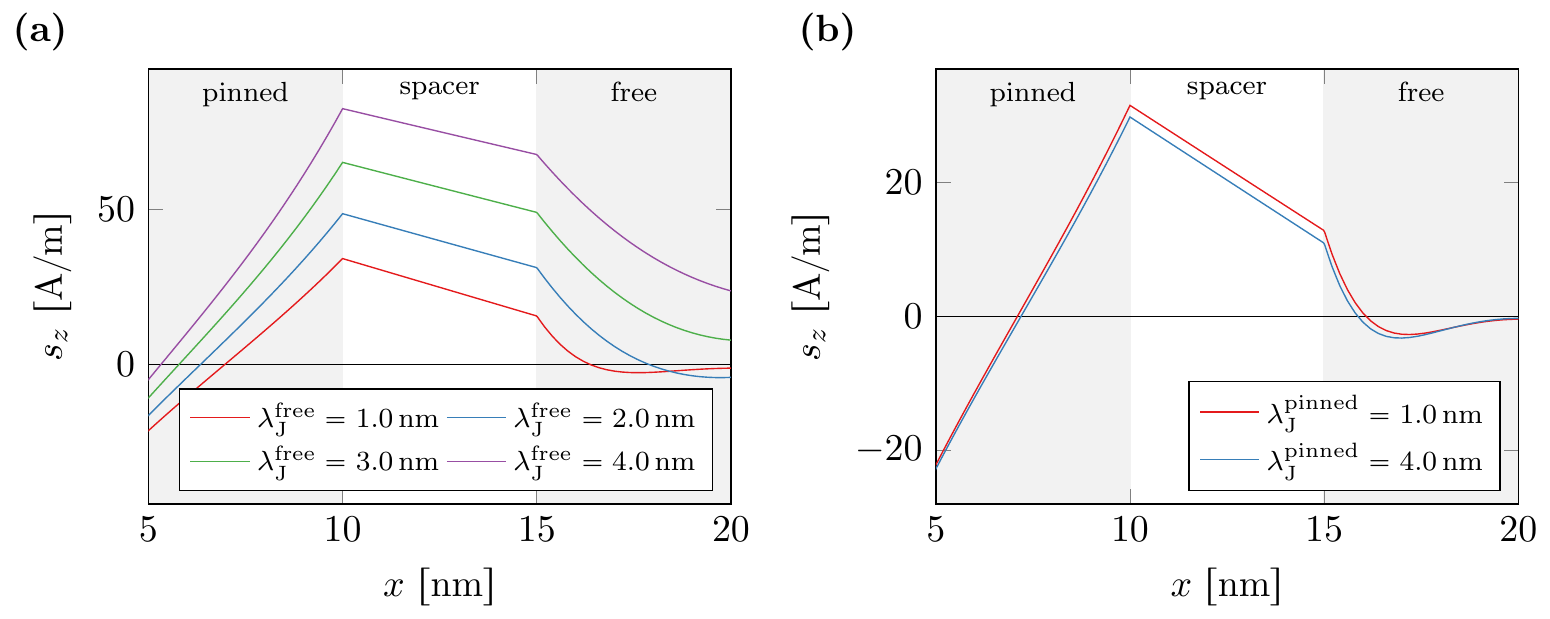}
  \caption{
    Space-resolved $z$-component of the spin-accumulation $s_z$ which is proportional to the field-like torque in the free layer.
    (a) $\lambdaj^\text{pinned} = \SI{1}{nm}$ and different $\lambdaj^\text{free}$.
    (b) $\lambdaj^\text{free} = \SI{1}{nm}$ and different $\lambdaj^\text{pinned}$.
  }
  \label{fig:s_z}
\end{figure}
For the magnetization configuration described above, namely homogeneous magnetization in $z$-direction in the pinned layer and homogeneous magnetization in $y$-direction in the free layer, the field-like torque strength $T_\text{field}$ is, apart from a constant prefactor, given by the $z$-component of the spin accumulation $s_z$.
This component is plotted in Fig.~\ref{fig:s_z} for varying characteristic lengths $\lambdaj$.
The spin accumulation, and thus the field-like torque, is always positive at the interface between free layer and spacer layer.
However, depending on $\lambdaj$ in the free and pinned layer the spin accumulation performs a rotation, which may lead to negative values of $s_z$ in parts of the free layer.
Comparing Fig.~\ref{fig:s_z}(a) and (b) reveals, that the influence of $\lambdaj$ in the free layer has a significantly larger impact on this behaviour.
A high $\lambdaj$ corresponds to a low $J$ and thus a low spin coupling of the itinerant electrons with the magnetization.
In this case, the behaviour of the spin accumulation in the free layer is similar to the behaviour in a nonmagnetic region, namely the spin accumulation decays as $e^{-x/\lambdasf}$.
In the case of low $\lambdaj$, the spin accumulation experiences a torque due to the magnetization as the magnetization experiences a torque due to the spin accumulation.
This torque explains the rotational behaviour of the spin accumulation and thus also the possibility of negative values for $s_z$.
Depending on the characteristics of this oscillation, the field-like torque becomes negative not only in parts of the free layer, but also in average.
For strongly exchange coupled systems this means that the overall field-like torque in the free layer can have a negative sign.

Besides the possible sign change of the field-like torque, it should be noted that, depending on $\lambdaj$, the field-like torque may become as large as or even larger than the damping-like torque.
This is interesting since the field-like torque is usually assumed to be much smaller than the damping-like torque and hence it is not considered to be relevant for applications.
This bevaviour occurs at high $\lambdaj$ in the free layer and low $\lambdaj$ in the pinned layer, e.g. $\lambdaj^\text{free} = \SI{4}{nm}$ and $\lambdaj^\text{pinned} = \SI{1}{nm}$ results in $T_\text{damp} = \SI{19456}{A/m}$ and $T_\text{field} = \SI{27933}{A/m}$.

\begin{figure}
  \includegraphics{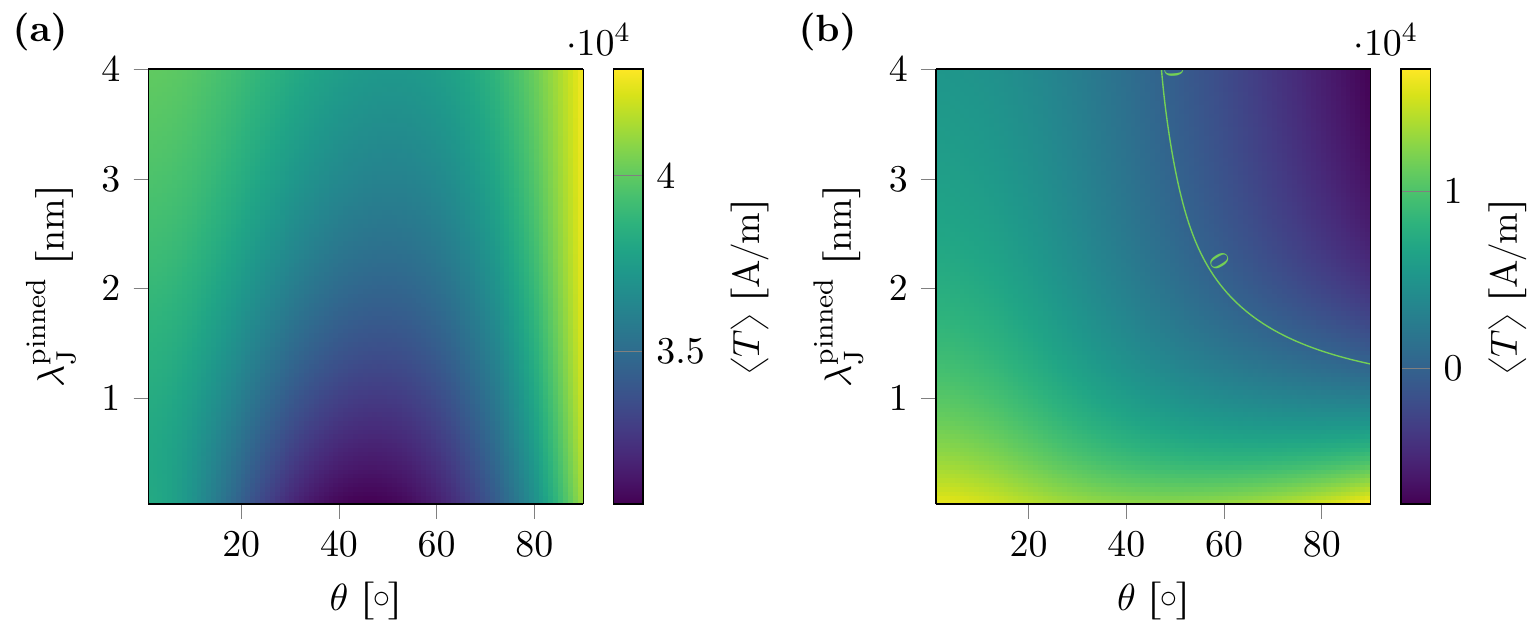}
  \caption{
    Damping-like and field-like torque for different exchange lengths in the pinned layer $\lambdaj^\text{pinned}$ and tilting angle between magnetization in free and pinned layer $\theta$.
    (a) Damping-like torque $T_\text{damp}$.
    (b) Field-like torque $T_\text{field}$.
  }
  \label{fig:tilt_ljp}
\end{figure}
Up to now the free layer was considered to be magnetized in $y$-direction and thus perpendicular to the pinned layer.
Fig.~\ref{fig:tilt_ljp} shows the averaged damping-like and field-like torques for different tilting angles $\theta$ of the magnetization in the pinned layer and free layer.
Here, $\theta = \SI{0}{\degree}$ means that the magnetization of the free layer points in $z$-direction like the pinned layer and $\theta = \SI{90}{\degree}$ means that the free-layer magnetization points in $y$-direction.
As for the preceding experiments, the damping-like torque does not significantly depend on the change of parameters.
However, the field-like torque is very sensitive to the tilting angle $\theta$.
For large $\lambdaj^\text{pinned}$ the field-like torque may even change its sign depending on $\theta$.

\begin{figure}
  \includegraphics{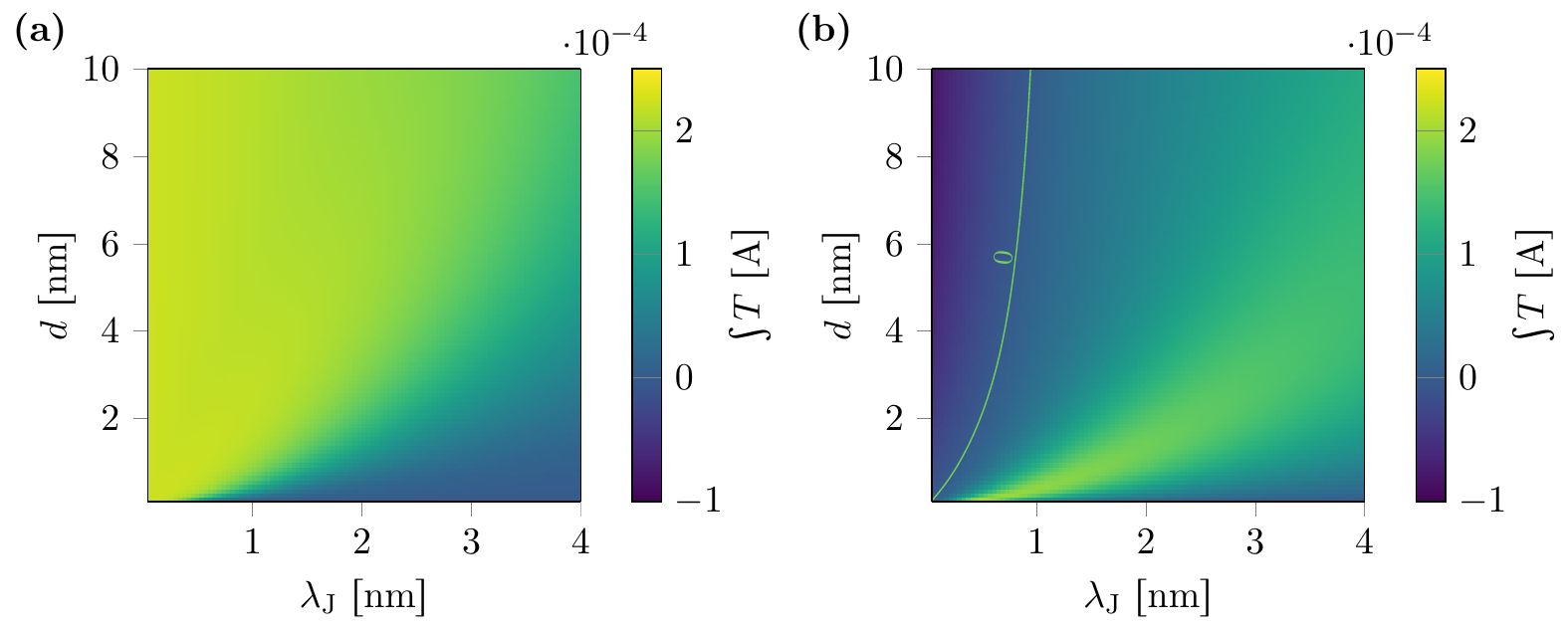}
  \caption{
    Damping-like and field-like torque for different thicknesses $d$ and spin-flip relaxation lengths $\lambdasf$ of the free layer.
    (a) Integrated damping-like torque $\int T_\text{damp} \dx$.
    (b) Integrated field-like torque $\int T_\text{field} \dx$.
  }
  \label{fig:ljf_d}
\end{figure}
In a last experiment, the interplay of the free-layer thickness $d$ and the exchange length $\lambdaj$ in the free layer is investigated.
Fig.~\ref{fig:ljf_d} shows the damping-like and field-like torque in the free layer.
In contrast to the previous experiments, the torque is integrated and not averaged over the free layer in order to give a proper measure of the overall torque for the different layer thicknesses.
The results can be explained with two different effects.
For very small free-layer thicknesses, well below the the characteristic length $\lambdaj$, the spin of the itinerant electrons is not completely transferred to the magnetization which results in a low overall torque.
The critical thickness decreases for increasing $\lambdaj$ which corresponds to low spin coupling.
This effect can be clearly seen in both torque terms, Fig.~\ref{fig:ljf_d}(a) and (b).
For the field-like torque, the rotation in the spin accumulation, as seen in the previous experiments, leads to low or even negative values for low $\lambdaj$.
Fig.~\ref{fig:ljf_d}(b) shows an approximately linear region of maximum field-like torque.
It should be noted that the field-like torque is of the same order of magnitude as the damping-like torque in the maximum region.

\section{Conclusion}
The damping-like and field-like torque in the free layer of a magnetic trilayer structure has been investigated with a spin-diffusion model.
The exchange coupling of the itinerant electrons with the magnetization has been found to have a major impact on the strength and sign of the field-like torque.
While the field-like torque is usually considered to be small compared to the damping-like torque, we show that for a weak exchange coupling in the free layer, the field-like torque can excel the damping-like torque.
On the other hand, for a very strong coupling the sign of the field-like torque may change.

\appendix
\section{Effective material parameter for finite leads}\label{sec:effective_leads}
\begin{figure}
  \centering
  \includegraphics{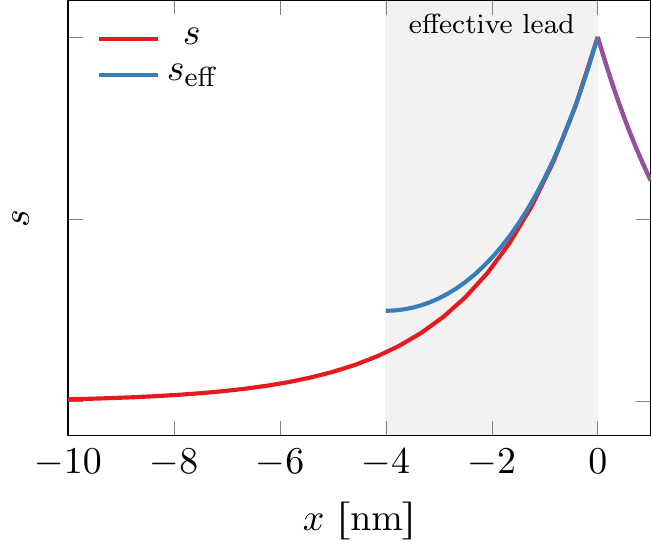}
  \caption{
    realistic spin accumulation $s$ for infinite leads and effective spin accumulation $s_\text{eff}$ for finite leads.
  }
  \label{fig:ap_s}
\end{figure}
In experiments the size of the leads is usually much larger than the spin diffusion length $\lambda_\text{sdl}$ in the lead material.
In order to retrieve accurate simulation results, the size of the leads has to be chosen accordingly, which adds a huge amount of additional degrees of freedom.
However, by choice of an appropriate effective spin-diffusion length, the effect of infinite decay can be perfectly modeled with finite leads.
Without loss of generality, we consider the interface between lead and magnetic region to be at $x = 0$, see Fig.~\ref{fig:ap_s}.
Assuming an infinite decay of the spin accumulation towards $-\infty$ yields
\begin{equation}
  s(x) = a e^{x / \lambdasf}
  \label{eq:ap_s}
\end{equation}
where $\lambdasf$ is the spin-flip relaxation length, which equals the spin diffusion length $\lambda_\text{sdl}$ in metal, and $a$ is constant determined by the solution of the model in the complete space.
In order to simulate the infinite leads with finite leads, we introduce an effective spin accumulation $\seff$ that may contain decaying and ascending contributions
\begin{equation}
  \seff(x) = a' e^{x / \lambdasf'} + b' e^{-x / \lambdasf'}.
  \label{eq:ap_s_eff}
\end{equation}
For the accurate solution of the spin accumulation in the magnetic material both, the spin accumulation $\seff$ and its first spatial derivative $s_\text{eff}'$, have to be correct at the lead--magnet interface
\begin{align}
  \seff(0) &= s(0),\\
  \seff'(0) &= s'(0).
\end{align}
We solve the spin diffusion equation with homogeneous Neumann boundary conditions.
Hence, additionally it holds
\begin{equation}
  \seff'(-d) = 0
\end{equation}
where $d$ is the finite thickness of the lead.
Inserting the definitions \eqref{eq:ap_s} and \eqref{eq:ap_s_eff} yields the system
\begin{align}
  a' + b' &= a \label{eq:ap_system1}\\
  \frac{a'}{\lambdasf'} - \frac{b'}{\lambdasf'} &= \frac{a}{\lambdasf} \label{eq:ap_system2}\\
  \frac{a'}{\lambdasf'} e^{-d / \lambdasf'} - \frac{b'}{\lambdasf'} e^{d / \lambdasf'} &= 0 \label{eq:ap_system3}.
\end{align}
From \eqref{eq:ap_system1} and \eqref{eq:ap_system2} we can derive
\begin{align}
  a' &= \frac{a}{2}\left(1 + \frac{\lambdasf'}{\lambdasf}\right)\\
  b' &= \frac{a}{2}\left(1 - \frac{\lambdasf'}{\lambdasf}\right).
\end{align}
Inserting into \eqref{eq:ap_system3} yields
\begin{equation}
  \left( 1 + \frac{\lambdasf'}{\lambdasf} \right) e^{-d / \lambdasf'}
  - \left( 1 - \frac{\lambdasf'}{\lambdasf} \right) e^{d / \lambdasf'}
  = 0.
  \label{eq:effective_lambdasf}
\end{equation}
Given the real diffusion length of the infinite lead $\lambdasf$ and a finite lead thickness $d$, this system can be solved for the effective diffusion length $\lambdasf'$.
This procedure is precise and only suffers from discretization errors.

\section*{Acknowledgements}
The financial support by
the Austrian Federal Ministry of Science, Research and Economy and the National Foundation for Research, Technology and Development
as well as
the Austrian Science Fund (FWF) under grant F4112 SFB ViCoM,
the Vienna Science and Technology Fund (WWTF) under grant MA14-44,
and MEXT Grant‐in‐Aids for Scientific Research (16H03853) of Japan
is gratefully acknowledged.

\bibliographystyle{ieeetr}
\bibliography{refs}

\end{document}